# Securing Data in Storage: A Review of Current Research

Paul Stanton

Department of Computer Science, University of Illinois at Urbana-Champaign

**ABSTRACT**

Protecting data from malicious computer users continues to grow in importance. Whether preventing unauthorized access to personal photographs, ensuring compliance with federal regulations, or ensuring the integrity of corporate secrets, all applications require increased security to protect data from talented intruders. Specifically, as more and more files are preserved on disk the requirement to provide secure storage has increased in importance. This paper presents a survey of techniques for securely storing data, including theoretical approaches, prototype systems, and existing systems currently available. Due to the wide variety of potential solutions available and the variety of techniques to arrive at a particular solution, it is important to review the entire field prior to selecting an implementation that satisfies particular requirements. This paper provides an overview of the prominent characteristics of several systems to provide a foundation for making an informed decision. Initially, the paper establishes a set of criteria for evaluating a storage solution based on confidentiality, integrity, availability, and performance. Then, using these criteria, the paper explains the relevant characteristics of select storage systems and provides a comparison of the major differences.

## 1. INTRODUCTION

With the proliferation of stored data in all environments, organizations face an increasing requirement to both temporarily and permanently retain information. The storage medium for housing this information becomes a prime target for attack by a malicious intruder. If an outsider can successfully penetrate the data storage, the intruder can potentially gain information that violates privacy, that discloses valuable secrets, or that prevents the access of legitimate users; the deleterious effects of such an attack are truly unquantifiable. If the organization takes no storage security measures, the data store becomes a lucrative single point of attack for an intruder. The avoidance of this obviously unfavorable condition has generated a detailed field of computer research. Universities have actively pursued options for securing stored information, and have consequently developed many potential schemes for ensuring information confidentiality, integrity, and availability without substantially degrading performance. See Figure 1 for a relational diagram of ongoing research.

A major problem associated with storing large amounts of data is how to properly weigh the costs and benefits associated with security measures. The most secure systems are so because of the increased measures to protect the data, but each additional measure comes with a cost in terms of both convenience and processing time. In order to effectively select the best security scheme, users must have an understanding of the primary security features available in the storage security community and then be able to quantifiably compare the systems. Developing an understanding of these aspects will help to motivate the direction for future research and assist the selection of the appropriate storage solution for a set of specific requirements.

The remainder of the paper is organized as follows. Section 2 provides a standardized set of criteria to evaluate secure storage systems. Section 3 provides a survey of eight storage systems. Section 4 provides a classification and comparison of the surveyed systems, and Section 5 concludes.

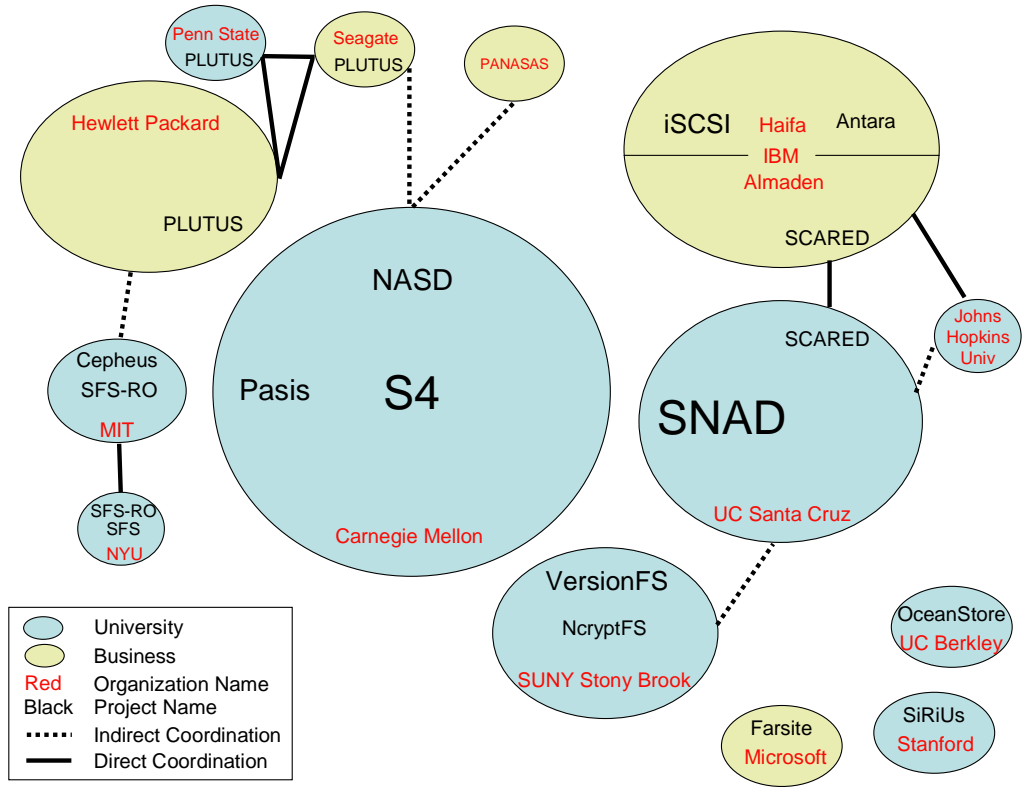

*Figure 1  Relational diagram of current research*

## 2. CRITERIA FOR EVALUATION

This section establishes a common set of criteria for evaluating a storage security system. There are many different ways to approach storage systems but for the purposes of establishing a common reference, confidentiality, integrity, availability, and performance have been selected. While this paper does not approach any criteria in exhaustive detail, it is necessary to describe the evaluation criteria prior to assessing the individual systems. Confidentiality, integrity, and availability are commonly referred to in the computer security arena, and performance was added to ensure systems achieve an appropriate balance between security and processing ability. Prior to discussing each aspect in more detail, it is important to understand that none of these attributes is mutually exclusive, and, in fact, to have a secure system all attributes must be satisfied.

2.1 Confidentiality

From a security perspective, ensuring confidentiality implies that no one has access to data unless specifically authorized. Different systems control this authorization process in various ways. The first step in authorizing access to information is to properly identify users via authentication. The storage system must define the means for a user to be properly identified prior to gaining access, and then having appropriately identified a user, the system must allow access to only specified data associated with that user. Proper authorization to access the storage system does not imply access to the entire system, in fact, the contrasting principle of least privilege is generally applied. Data owners must, however, have a method for allowing others to access information when appropriate via a delegation of authorization scheme.

In addition to managing authorization to data, confidentiality also implies that the system must encrypt data to prevent information attacks. Therefore, the system must require either users or servers to apply cryptographic keys. The differing design decisions between user managed and sever managed keys have had significant impact on the overall storage technique. In order to share information, multiple users must have access to the appropriate keys – whether a centralized group server hands out keys or individual file owners provide the keys to additional users, the effects on performance and user convenience must be analyzed. The discussion of key management in this paper is not intended to detail cryptography, but an understanding of how keys are distributed and applied is essential to understanding the larger system. Since the cryptographic operations are often the most computationally expensive aspect of accessing securely stored data, understanding how a particular system manages keys is appropriate.

An additional critical discussion concerning key management involves how keys are revoked. Once an owner or administrator determines to revoke a particular user's access to data, the keys that the user had must no longer allow access to the system, or if they do they must not allow access to future versions of the files. The cost associated with revoking a user manifests itself in the re-encryption effort required to secure confidentiality. It is not possible to physically revoke a user's keys to prevent that user's ability to perform operations since copies could have been produced, so the system must render all keys of a revoked user obsolete and re-encrypt all of the data with a new key. A resulting argument then turns, once again, to a tradeoff between security and performance. There are two primary methods for securing the data after key revocation: lazy or aggressive revocation. When using lazy revocation the system does not re-encrypt the data that the revoked user previously had authorization to access until the next valid user attempts to access the file. This essentially defrays the cost over time, but it leaves data vulnerable to the revoked user for an unspecified period of time. By contrast, aggressive revocation immediately re-encrypts all files that the revoked user could potentially access. Once re-encrypted, new keys must be distributed to all personnel who are affected by the changed encryption (adding additional weight to the key distribution scheme); clearly this option requires time. Lazy re-encryption sacrifices a measure of security to save time while aggressive revocation sacrifices time to improve security.

## 2.2 Integrity

Integrity is a broadly based topic that includes maintaining data consistency in the face of both accidental and malicious attacks on data. For the purposes of this paper, the scope of the integrity analysis is limited to the methods used to prevent malicious alteration or destruction of information. The resulting expectation is that when a user accesses stored information, no data has been subjected to unauthorized modification. Many systems enforce integrity by ensuring that data comes from the expected source. For stored data, the discussion of integrity implies that files have not been changed on the disk.

Integrity enforcement procedures fall into two categories: data modification prevention and data modification detection. Similar to confidentiality, modification prevention requires users to receive authorization prior to changing files and requires that files are only changed in an approved manner. Integrity varies from confidentiality in that confidentiality is only worried about whether or not data has been compromised, whereas integrity includes ensuring the correctness of the data. Detection schemes generally assume that attacks are inevitable and that there must be suitable ways to assess any damage done, recover from the damage, and apply lessons learned to future prevention mechanisms.

## 2.3 Availability

The paper considers availability in terms of time, space, and representation. Information needs to be available to an authorized user within an acceptable time period, without monopolizing the available storage space, and in an understandable representation. A system can not allow an adversary to prevent authorized access to information via a denial of service attack.

It is important to note that the goals of availability conflict to a degree with those of confidentiality; the two must be considered within the security domain.

## 2.4 Performance

The level of security and the system performance often conflict. In order to provide the requisite layers of security to avoid harmful attacks, the system performance suffers. The two goals of an efficient system and a secure environment intrinsically conflict. Each additional security measure requires computationally expensive processing that detracts from the system's ability to perform other operations; all security measures are overhead for the system. Each of the evaluated storage techniques attempts to minimize the performance cost associated with the particular measures of the system.

The most dominant performance cost is associated with encryption due to its computationally expensive nature. The two fundamentally different approaches to storage security, encrypt-on-wire and encrypt-on-disk, place the burden of encryption on different aspects of the system. Riedel et al [25] provide a detailed explanation of the tradeoffs between the two.

## 3. SURVEY

The following survey briefly describes several storage security approaches using confidentiality, integrity, availability, and performance as a framework. Sections 3.1-3.3 refer to encrypt on wire systems and Sections 3.4-3.8 refer to encrypt on disk systems.

**3.1 NASD** – **Network Attached Secure Disks**.

In traditional distributed file systems, a client wanting to access data must make a request to the file server. The server then must verify the client's authorization and distribute the file if the appropriate criteria are met. Since the server must interact with every file access request for every client, the server can quickly become a bottleneck. NASD's primary goal is to relieve the server bottleneck by interacting with a user one time providing a "capability key." With the capability key the user can access the appropriate disk(s) directly without any further server interaction. The disks themselves must be "intelligent" such that they possess enough internal ability to process the capability key and handle file access requests directly [10,11].

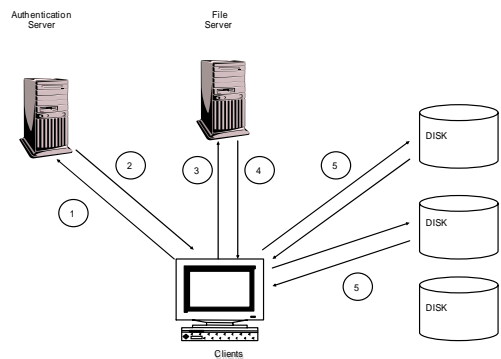

1. Client requests authentication
2. Server responds
3. Client request to file server
4. File server responds with capability object
5. Request directly to drive
6. Response from disk

*Figure 2 NASD*

Confidentiality. There are two servers in the NASD design, one to provide authentication and then the actual file server. NASD does not specify the authentication scheme and recommends using any existing method similar to Kerberos. Upon receipt of authentication, a user sends a request to the file server. The server verifies the authenticity of the request and then provides the user with a capability key that corresponds to the user's rights for file access. After obtaining the capability key, a user can communicate directly with the data disk for all future access requests during a given session.

The capability object is the critical aspect pertaining to both the confidentiality and integrity of the system. A file manager agreeing to a client's access request privately sends a capability token and a capability key to the client; together these form a capability object. The token contains the access rights being granted for the request and the key is a message authentication code (MAC) consisting of the capabilities and a secret key shared between the file server and the actual disk drive. Clients can then make a direct request to a NASD drive by providing the capability object. The drive then uses the secret key that it shares with the file server to interpret the capability token to verify the user's access rights and service the request. Since the MAC can only be interpreted using the drive/server shared secret key, any modifications to the arguments or false arguments will result in a denied request.

Integrity. The novel concept associated with NASD is placing part of the data integrity requirement on the disks themselves. The "intelligent" disks interpret the capabilities objects, encrypt data, and transmit results to clients. To ensure integrity on the client end, the disk uses the same hash MAC combination that allowed it to authorize a client access to encrypt and send the data to the client. The client can then verify the integrity of the transmission during the decryption process.

Availability. The fact that NASD allows direct access to the disk promotes scalability; the system throughput scales linearly with the number of clients and disks. However, since the file server must be trusted to initially provide capability keys, the server presents a single point of attack. If the server becomes compromised there is no way to prevent a denial of service attack.

Performance. A motivating factor for using NASD is the ability to scale bandwidth linearly with the number of disks in the system, however, these benefits are partially offset by the cost of cryptography. A large performance problem associated with NASD is the dual cost of cryptographic operations incurred due to the encrypt-on-wire scheme. Every data transmission must be encrypted prior to being sent and then decrypted at its destination from disk to client or from client to disk. In an attempt to reduce the performance penalty, NASD uses a "hash and MAC" cryptographic approach instead of a standard MAC. In a traditional MAC algorithm, a client's secret key is used throughout the computation. In contrast, hash and MAC uses the raw data from the file to pre-compute a series of message digests that are generic for the given file. Hash and MAC then applies a client's secret key to the message digests only as a client requests a file. The result is that the secret key is only required for a small subset of the overall computation, thus significantly decreasing latency associated with on-the-fly cryptography. Experiments demonstrated that the latency for using cryptographic operations was bounded by a 20% increase in the time to service a request when compared to a request with no cryptography [10].

**3.2 PASIS – Survivable Storage**

PASIS is a survivable storage system designed to address problems associated with compromised servers; the system assumes that compromised servers will exist and therefore addresses how to protect data in such an environment. PASIS employs a threshold scheme to distribute trust among storage nodes to prevent data security breaches even when faced with a compromised server. The threshold scheme encodes, replicates, and divides information such that the pieces of data are stored in different locations. In order to make the data disbursement transparent to users, PASIS requires a client side "agent" to interpret user-level commands and the associated responses from the various PASIS servers connected to the storage nodes [8,33].

Confidentiality. PASIS strives to prevent data compromise by storing elements of a file in different locations so that a single compromised server cannot disclose any relevant information. Instead of cryptography to ensure confidentiality, PASIS uses a ramp $p$-$m$-$n$ threshold scheme that divides data into $n$ shares such that any $m$ of the shares can reconstruct the original data, but fewer than $p$ shares reveals no information about the original. (Cryptography and a threshold scheme can be combined to increase the level of protection, however, any cryptography would have to be layered on top of PASIS). As long as less than $p$ shares are ever visible to an intruder no information will be compromised.

Integrity. PASIS helps provide data integrity by not relying on any specific set of PASIS servers to provide the required $m$ shares of the data. Since having any set of $m$ shares allows the client agent to reconstruct the original data, those $m$ shares can come from any of the various servers in the network. In order to prevent data integrity an intruder must compromise the $m$ servers servicing the request and alter the data. If the client agent does not receive the requisite $m$ shares or is unable to recreate the original file due to malicious intervention, the request is rebroadcast.

Availability. Similar to the argument made for the PASIS integrity enhancements, the system's requirement for only $m$ shares to retrieve data increases data availability in the face of failed servers. The number of servers required in the "surviving" subset of servers has an upper bound of $m$, such that ($n$-$m$) servers can be compromised or unavailable and the system will still successfully service the request. Similarly, with a write operation the system administrator can determine the number of shares required to accept the write. At least $m$ shares must be successfully written, but any number between $n$ and $m$ will provide the correct data. Clearly, the more successfully written shares will provide for greater availability on subsequent file access requests.

     Even though PASIS provides for increased data availability for single users, it does not directly address concurrent access or concurrent modifications to files. This implies that some additional mechanism must be layered on top of PASIS to guarantee atomicity, which in turn implies potential message passing overhead or latency when multiple users access the same file simultaneously.

Performance. In relative comparison to a traditional distributed file system, PASIS is hindered by an increased number of message passes to receive the same information. A traditional system sends a request to a single server, whereas PASIS must broadcast requests to at least $m$ servers and then combine the resultant messages on the client machine. It is difficult to quantify the overhead associated with PASIS because there are great performance trade-offs associated with selecting different values for $n$-$m$-$p$. The values, however, can be customized for a particular file. For example, increasing the value of $n$ increases the likelihood that $m$ shares will be available, but it also means that more shares of the file are stored, thus increasing the potential for theft. The benefit of this flexibility allows users to select appropriate values for each file they store in the system. During their research, the PASIS designers discovered a significant performance cost for accessing small files, but a negligible penalty for large files [33].

**3.3 S4 – Self Securing Storage**

S4 is a self-securing storage medium that introduces a new aspect to storage security: the disks do not trust even the host machine operating system. S4 treats all requests as suspect. The driving security motivation for the system has been to negate the effects of a clever intruder who is able to successfully penetrate the operating system and disguise any adversarial efforts. The disks themselves in S4 require a small set of fundamental operations for managing a file system and therefore have an embedded instruction set for internally versioning and auditing all data and metadata. S4 uses a daemon on the client machine to service file access requests as remote procedure calls and then translate them into S4-specific requests to make the system transparent to users. [26, 13, 28, 27].

Confidentiality. S4 does not provide any method for authentication, but rather assumes that the self-securing disks will be used in conjunction with a file server that uses one of many standard authentication protocols. All access requests properly sent to S4 will be serviced without any additional verification.

Integrity. The focus of the system then is not designed for confidentiality, but rather for ensuring data integrity. An underlying theme for the design of S4 is that confidentiality will eventually become breached in any system. Analyzing the history of breached systems and studying the success of intruders, the researchers at Carnegie Mellon University assumed compromised servers as an unfortunate reality. They made the primary focus for their storage research preventing an intruder from damaging the stored data and capturing the intruder's actions through detection and diagnosis.

S4 uses a comprehensive versioning protocol that creates a new version of file metadata for every file access. S4 establishes a detection window during which all data and metadata are comprehensively versioned. In addition to versioning obvious write activity that changes the data itself, each file access also results in an update to the file's log so that any improper access can later be questioned. The period of the detection window is largely defined by the amount of space available for keeping the overhead of multiple versions and has proven to be approximately two weeks in research [26]. Considerable research has resulted in a combination of a journal-based structure and multi-version B-trees to efficiently keep all of the requisite information in a space efficient manner.

If at any time during the detection window an intrusion occurs, the system can guarantee the integrity of all data up to the point of the intrusion. Additionally, after verifying the access logs for individual files, files can be safely recovered if no anomalies exist. Files that have questionable accesses in the post-intrusion period may result in lost data, but the user can still be guaranteed not to receive the tainted file. This is a critical aspect of S4, other systems may allow a user to unknowingly access a file modified during an intrusion, but S4 will prevent such an integrity violation. The system administrator establishes the detection window, and once set, no user can prevent any access to data from being versioned. This precludes an intruder from altering a file in an undetectable fashion.

Availability. Through comprehensive versioning, self-securing storage inherently ensures data is available to users. Immediately after the system administrator detects an intrusion, all of the files can be reliably restored to the last access prior to the intrusion. Legitimate changes made after the intrusion but prior to its detection may result in lost data, but the lost data can be minimized. S4 includes an inherent intrusion detection scheme that can analyze the access history of files that the system administrator deems appropriate. If, for instance, an intruder attempts to change the contents of a password file which the administrator is "watching," the intrusion can be detected immediately. By minimizing the deleterious effects of an intrusion, the system naturally increases data availability.

Performance. Considering the costs of maintaining versions for every file access, the designers searched for both space and time efficient means for achieving comprehensive versioning. They use journal-based metadata and multi-version binary trees to meet their objectives. The system has proven to operate at a comparable speed with traditional NFS for current-file lookup. For back-in-time access, the lookup time is dependent on the number of versions which must be traversed. However, this can be bounded by a system administrator's determination to checkpoint the file system.

**3.4 CFS Cryptographic File System**
      CFS is archaic, but it is relevant to discuss CFS because it formed a theoretical stepping stone for other researchers to establish system design goals. A primary motivation for CFS was to eliminate the requirement for user or system-level cryptography and instead place the requirement in the file system. Manual or application based cryptographic operations were either error-prone or incompatible with one another. System level solutions lacked portability caused by embedded encryption techniques, they lacked compatibility due to specialized server authentication software, or they left potential security holes where data was temporarily stored as clear text. CFS proposed pushing all file encryption into the client file system. It used a /crypto mount point in Unix to mask CFS specific operations allowing the file system to treat encrypted files like any others. The system was designed for a local, not distributed, use, therefore an individual user must physically "hand-out" the cryptographic keys for each file [3].

Confidentiality. The only method for controlling authorization is the file owner's selectivity with passing requisite keys out to other users. It is the user's responsibility to ensure that the keys are distributed in a secure manner to only the correct and intended personnel. The method, while not scalable, does not rely on a trusted server. The owner of a file encrypts it with a symmetric key prior to writing it to the file system. Neither the file system, nor any users, ever has access to the clear text data. There are no special provisions for ensuring that the data is encrypted on the wire, but one can operate under the assumption that if the file leaves the client machine in encrypted form, is never modified by the server or the file system, and then is transmitted to another authorized client with the proper key that the information has always been secure.

CFS mounts a virtual file system (/crypt) to a standard Unix files system, and then directs all system calls related to encrypted files through the mount point. Users create directories under the /crypt mount point with an associated key which will then be used to encrypt all data stored within the directory.

Integrity. CFS converts standard NFS system calls into CFS specific calls using a daemon on the client machine. The daemon then issues RPCs to the file server after the client establishes a proper connection with the server. Any attempt to send a RPC directly to the file server, thus bypassing the CFS daemon, will be denied because of a requirement for all RPCs from a client to have been generated from a privileged port. This helps to prevent any malicious user from having access to modify files, but there is no direct mechanism to provide additional integrity protection. CFS relies on the assumption that the file server never has access to unencrypted data to ensure data integrity.

Availability. While all files remain encrypted on the file server, there is no mechanism to prevent an adversary from denying a legitimate user from accessing files if the server is compromised. The system does, however, use the underlying file system's sharing semantics to allow concurrent access to multiple users. Once keys are properly distributed, CFS provides comparable standard use availability to Unix.

Performance. CFS runs at user level and interfaces with the underlying file system via remote procedure calls. This implies that there is potential for significant context switch overhead in addition to the added cost of DES cryptographic operations. CFS proved to be up to 4 times slower than standard NFS for reading and writing large files, twice as slow for creating small files, and 30% slower for a mix of "standard" operations.

### 3.5 SFS-RO – Secure File System – Read Only

SFS-RO relies on self-certifying path names to provide high availability to read-only data in a distributed environment. SFS-RO uses some of the concepts from its SFS predecessor, but strives for better performance by providing read-only data that does not require any server-based cryptographic operations. The concept is to still ensure data integrity while producing multiple copies of read-only material; traditionally such copying resulted in a degradation of security [7, 17].

Confidentiality. SFS-RO relies on a mutual authentication protocol between the users and the server, performed via self-certifying pathnames that have the public key for a file embedded in them. The creator of the file has the ability to assign the key, therefore offering a wide range of cryptographic options.

To properly encrypt files, an administrator bundles the contents of the file system into a database that is signed with a digital signature containing the private portion of an asymmetric key. Once signed, the database can be replicated and distributed to many untrusted machines without the threat of compromise. In order to access the files, a user must provide the location of the storage server (either a DNS hostname or IP address) and a *HostID*. The HostID is a cryptographic hash of the server location and the public portion of the asymmetric key with which the file creator encrypted the database. The database creator must provide the public key to all potential users separate from the SFS-RO system.

      Once granted permission to the files via the mutual authentication, the users can then access files by providing the appropriate *handle*, comprised of a cryptographic hash of the file's blocks. Groups of handles are recursively hashed and stored in hash trees such that the handle to the root inode provides the ability to verify the content of individual file blocks, reducing the number of handles required throughout the system. Knowing the handle of the root inode provides a client with the ability to verify the contents of a given file by recursively checking the hashes.

<u>Integrity.</u> SFS-RO relies on three critical elements: the SFS database generator, the SFS read-only server daemon, and the client. Traditional directories are converted to a database and digitally signed in a secure client environment. This database is then distributed to any number of servers that all run the SFS-RO server daemon. The server daemon simply receives requests from clients to look up and return data. The SFS client runs on a client machine and is a conduit between a standard file system protocol (like NFS) and the server. Upon receipt of a file transmission, it converts the SFS-RO database "chunks" into traditional inodes and blocks that a typical file system would expect to see. Additionally the client must posses a private key to verify the digital signature on data passed from the server since SFS-RO does not trust the server. This verification process ensures data integrity.

      SFS-RO also uses a timestamp protocol to help detect integrity violations. When a user creates a database, the time is recorded. Additionally, the creator must establish a no-later-than time to resign the database so that the time has an upper and lower bound. Users of the files maintain a record of the current timestamp which they compare against all data that they receive to prevent a rollback attack.

<u>Availability.</u> One of the primary goals of SFS-RO is to extend access to read only data in a global environment. To accomplish this, a file system creator can copy the securely generated database onto any server that is running the SFS daemon. The result is a system that scales to the number of servers multiplied by the number of connections per server. Since the system is designed for multiple copies of read-only material to be distributed amongst multiple machines, it is logical to deduce that destruction of one server will not affect the availability of the file.

<u>Performance.</u> All cryptographic operations are performed on client machines; the creator establishes the database in a secure non-networked environment and users receive encrypted data which they must decrypt on their client machines. The cryptographic operations proved to be the most costly aspect in comparison to traditional NFS. For small files, SFS-RO was twice as slow as NFS with the primary additional cost due to timestamp verification (to ensure integrity). SFS-RO incurs additional latency when compared to NFS because NFS is run in the kernel while SFS-RO must rely on system calls. In larger files where the proportion of system calls to data passed is smaller, the slow down was approximately 30% which, while much improved over small files, is still a considerable performance penalty. SFS-RO began to perform favorably with very large files (40MB), proving 4% faster than NFS.

### 3.6 SNAD – Secure Network Attached Disks

Secure Network Attached Disks are designed to prevent any unauthorized personnel from accessing stored files by encrypting all data, and only allowing decryption on a client machine. This eliminates the potential threat posed by compromising the system administrator's access rights or that posed by physically capturing a disk. The individual drives lack sufficient information to decrypt any data themselves, and rely instead on a key management scheme that provides an authorized user with sufficient keys to decrypt files on the remote client machine. [5, 20, 21, 24]

<u>Confidentiality.</u> The critical functionality behind SNAD lies in the lockbox mechanism for storing keys. Each file consists of variably sized secure data objects which are all individually encrypted with a symmetric object key. Within the file's metadata is a pointer to a key object for that file which, itself, can be considered a file. Within the key object's metadata there are fields for a unique key file ID, the user ID for the creator of the file, and a digital signature from the last user to modify the file. The digital signature of the last modifier of the file is provided to assure all other users that the key object itself has not been modified (any authorized user can verify that the key object signature matches with someone authorized to write to the file). The key object file itself consists of tuples that coincide with legitimate users for the original file. Each tuple contains a user ID field, the object symmetric key to access the secure data objects, and a listing of whether or not the user has permission to write to the key object (coinciding with permission to write to the original file). The object key is itself encrypted with the public portion of a user's asymmetric key so that the only way to decrypt the object key is with the user's private key on the user's client machine. This prevents any intruder from ever being able to access the symmetric key that actually encrypted the stored data.

In addition to the key objects, SNAD manages authorized users by maintaining a certificate object with tuples containing valid user IDs, the user's public key, a hashed message authentication code key to provide and verify user digital signatures, and a timestamp the system updates whenever the user performs a write operation to prevent replay attacks.

<u>Integrity.</u> To provide integrity enhancement, SNAD stores a non-linear checksum of the original data along with the encrypted data so that a user can verify that the file has not been maliciously changed during storage. The checksum is updated whenever an authorized user makes a modification to the file. Users can also verify the integrity of writes by analyzing the file's key object file metadata and checking the digital signature provided.

<u>Availability.</u> Since the lock-box of keys is a critical aspect of SNAD, it must be available for users to access files. Unfortunately, the lockbox is stored on a single trusted server which presents a single point of attack for an adversary. If the lockbox server is compromised, there is no way to prevent a denial of service attack. Additionally, the system does not have a specified key revocation policy and leaves the decision and implementation of active or lazy revocation to the file owner.

Performance. The computationally expensive encryption and decryption tasks are performed on the client machines and avoid any potential bottleneck at the server. The decryption process, however, can still be very slow even when performed on the relatively faster client machines. To offer a user options, the designers of SNAD have implemented three separate schemes for providing digital signatures that trade security for performance. A file's creator can determine the granularity with which to verify the digital signature, the finer the granularity the more security provided and vice versa. The developers have proven through empirical studies that the digital signature process is by far the most costly with SNAD [5]. The most secure method where digital signatures are provided with each block write and verified with each block read is not suitable for standard use. Only the least secure of the three options (verifying the digital signature based on a hashed MAC instead of a public key) proved to be comparable to a system without security [21].

**3.7 PLUTUS**

PLUTUS is another lockbox scheme that operates similarly to SNAD, but the primary goal of PLUTUS is to provide highly scalable key management while providing file owners with direct control over authorizing access to their files. All data is encrypted on the disk with the cryptographic and key management operations performed by the clients to alleviate server cryptographic overhead. Users can customize security policies and authentication mechanisms for their own files using the client-based key distribution scheme. This places the responsibility for key management on the user, forcing the file owner to ensure proper secure distribution of keys to those they wish to authorize access. A prospective user must contact the owner in order to get the appropriate key. Riedel et al [25] argue that this task can be performed with an acceptable cost to user convenience. [16]

Confidentiality. To provide the liberty of owner customization while ensuring confidentiality, PLUTUS relies on an intricate lockbox scheme with multiple levels of keys. At the data level, PLUTUS uses a block structure to encrypt each individual block with a unique symmetric key. These block keys are then encrypted within a lockbox accessed via a file-lockbox key common to all files within a filegroup. The filegroup owner creates the file-lockbox key when the file is created and then distributes it to all users. PLUTUS uses an asymmetric file-verify key or a file-sign key protocol to differentiate between readers and writers respectively (see Figure 3). These keys are used to sign or verify a cryptographic hash of the file block contents to provide integrity. Upon requesting a file, the server passes the encrypted lockbox and encrypted block contents to the user. The user then "unlocks" the lockbox with the file lockbox key and decrypts each block with its respective file block key.

The proliferation of keys and the use of filegroups in PLUTUS complicate the key revocation scheme. When multiple files across the file system (related only by access rights, not like a traditional directory) are encrypted with the same key, key revocation could cause mass re-encryption and key management problems. However, the designers have implemented a clever key rotation scheme that minimizes the effects. PLUTUS uses lazy revocation such that a revoked user can still read files that were accessible at the time of revocation. A problem arises due to the use of filegroups because upon re-encrypting a file, different files within the same group will require different keys. Since a primary motivation for using filegroups in the first place was to minimize the number of keys, Kallahalla et al [16] designed a rotation scheme that ensures that the new encryption key is related to the keys for all files in the filegroup. The system performs the re-encryption with the latest filegroup keys, but all valid users can generate previous versions from the latest key. The result is that all valid users can "regenerate" the proper key for a given file if they have the latest filegroup key. The new filegroup key is only disseminated to currently valid users such that revoked users.

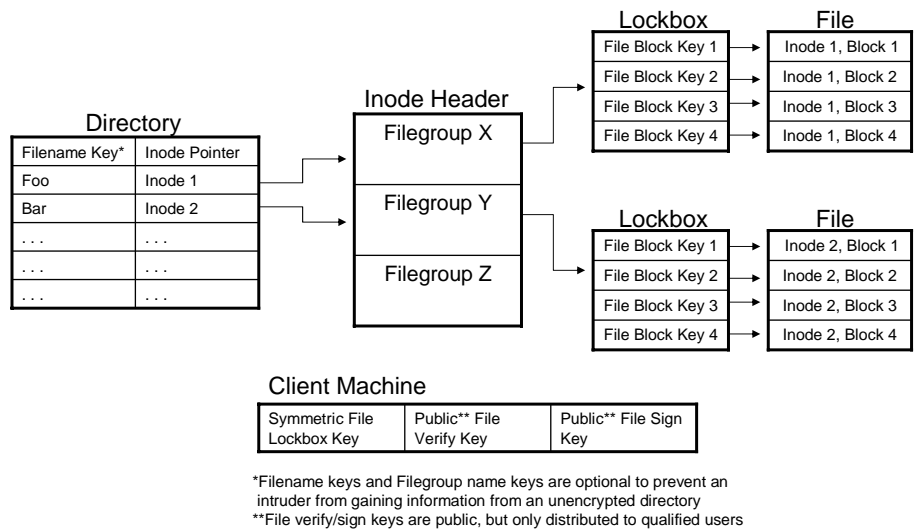

*Figure 3 PLUTUS Key System*

Integrity. PLUTUS does not trust the file server and cannot, therefore, rely on it to distinguish between writers and readers. Instead it uses two types of keys, file-sign keys and file-verify keys, to make the respective determination. Upon attempting to read or write, the user verifies the digital signature and hashed contents of the file with these keys. If the user obtains unexpected results, the user can determine that the file has been illegally modified.

Availability. The entire design of PLUTUS is intended to provide scalability. Placing the key management responsibility on the clients instead of on a trusted server prevents a server bottleneck due to computationally expensive cryptographic operations. PLUTUS relies heavily on filegroups to limit the number of cryptographic keys. Filegroups consist of all files with identical sharing attributes and can, therefore, be protected using the same key. This allows users with filegroup privileges to access a file within the group even if the owner is not on-line, avoiding the requirement for a user to contact the owner directly to get the key with every file access. The filegroup concept does not rely on a hierarchical structure so that the grouping is strictly a product of the associated files' permission attributes.

PLUTUS does provide much greater scalability and a clever key rotation process to minimize key management responsibilities associated with key revocation, but the system still requires the file's owner to provide a copy of the file's symmetric key to each user. Filegroups help to minimize the file owner-user communication, but they do not eliminate the original responsibility for the owner to distribute the keys.

Performance. The designers of PLUTUS used OpenAFS to construct their system. The performance of the file system itself is comparable to the unmodified OpenAFS system, because the server does not have to perform any extraneous operations during file access. However the cost to the client system where cryptographic operations are performed demonstrated to be 1.4 times slower than SFS. The authors present a credible argument to justify this decrease in performance, because they used worst-case scenarios to derive their figures. This comparison only takes into account a single file read and write combination which does not take advantage of the design enhancement offered by PLUTUS. PLUTUS is designed for scalability and to relieve server bottleneck which may arise with multiple file access requests in standard SFS. Kallahalla et al [16] reason that the average latency for protracted use would favor the use of PLUTUS.

**3.8 SiRiUS – Securing Remote Untrusted Storage**
SiRiUS is designed to provide its own cryptographic read-write file access on top of any existing untrusted networked file system (NFS, CIFS, Yahoo, etc.). Via a software daemon, the system intercepts all file access system calls and converts them accordingly. The concept is to be able to establish a secure file sharing environment without significantly modifying the performance of an existing network storage medium. SiRiUS can provide security to an existing system without requiring any hardware modifications; the developers view the system as a "stop-gap" measure to provide additional security to existing systems. Often times, organizations cannot afford to upgrade their current systems and must continue to operate with limited security until which time the option to upgrade security measures becomes available; SiRiUS can provide an interim solution [12].

Confidentiality. All files are encrypted in a secure environment prior to being stored on the server, such that neither the server nor the server administrator ever has access to unencrypted data. Additionally, the computationally costly encryption operations are performed on the relatively lightly loaded client machine, and the fact that the data is already encrypted obviates any requirement to establish a secure channel to send the file to the server. Each file owner maintains a master encryption key (MEK) and a master signing key (MSK). Each file has a unique symmetric file encryption key (FEK) provided to all users and a file signing key (FSK) provided only to authorized writers to the file. The system provides a "freshness guarantee" by maintaining a metadata freshness file for each directory.

All files are separated into two parts: an md-file metadata file and a d-file data file. The metadata file contains a block for the file owner's MEK, a block for every valid user's FEK (and FSK if authorized to write to the file), and a block with a hash of the metadata file's contents signed with the owner's MSK. If the owner or a user has a key maintained in the file's metadata, that person can decrypt the file. User key revocation is quick and efficient; the file owner removes the revoked user's key block from the metadata file, creates a new FEK, re-encrypts the file with the new key, and then updates the remaining users' key blocks with the new FEK. The result is immediate revocation.

Integrity. In addition to added measures, SiRiUS keeps certain file system specific metadata unencrypted so that the file system can perform standard integrity checking operations. SiRiUS keeps all access control information encrypted with the file data. This facilitates using the legacy file system's standard backup procedures – if the system must recover from a crash, all of the needed access information is already available with the file. SiRiUS uses the "freshness guarantee" to ensure that users have the most current version of a file preventing a rollback attack. At a user-designated interval, the user timestamps the metadata freshness file.

Availability. The design decision to make no modifications to the underlying file server prevents SiRiUS from defending against denial of service attacks; an attacker could conceivably compromise the server and delete all files. SiRiUS has no ability to intervene in such a circumstance and, therefore, requires users to backup their own files on multiple servers to limit the effects of such an attack.

In order to add users, the perspective reader/writer of a file must send a public key to the file owner who will then use the public portion of the MEK to encrypt that key and add it to the files metadata. Once the new user's key is added to the metadata, the user has access to the file. The key passing mechanism is not addressed in the system. As previously addressed, SiRiUS supports active key revocation such that once a user's access rights have been revoked, the user no longer has any form of access to the file via the freshness guarantee. SiRiUS attempts to improve on other secure networked file system designs by allowing fine grained file access while providing the ability for a file owner to grant read-only or read-write access to shared files. Other systems either allow access to entire directories or cannot distinguish between readers and writers.

Performance. The first time a file is accessed, the file system must return the associated metadata file as well as the original (to support appropriate authorization checking). The metadata file is then cached to prevent the overhead of looking up and sending the metadata file on subsequent retrievals for the same file. Additionally, many SiRiUS file system calls require checking the freshness file resulting in increased network traffic and additional file I/O for each. 63% of the cost associated with using SiRiUS for a 1MB data read is due to the decryption cost. Similarly, 40% of the cost for writing a 1MB file is due to signature generation.

# 4. CLASSIFICATION

| Storage System | Encryption Location | Trusted Server | Key Revocation Policy | Confidentiality Measures | Integrity Policy | Availability Policy | Estimated Performance Overhead |
|---|---|---|---|---|---|---|---|
| **NASD** | Wire | Yes | Active by using timestamps, key issued one time | Capability keys, separate authentication server | Hash MAC checksums to send data, not secure on disk | Scalable to many users, subject to DOS | 20% increase over system with no security |
| **PASIS** | Wire | No | N/A | *p-m-n* threshold scheme | Data dispersed across storage nodes such that > *p* must be compromised | Only *m* shares must be available to recreate the original | Significant overhead for small files, negligible overhead for large files |
| **S4** | Wire | No | N/A | N/A | Detection scheme: Comprehensive versioning | Intrusion detection and diagnosis to provide "recent" version | Comparable to NFS |
| **CFS** | Disk | Yes | Manual active revocation | Keys handed out to users by the file owner | Privileged port combined w/ encryption | Limited to manual distribution of keys/ DOS if server compromised | 30% slower than NFS for standard workload |
| **SFS-RO** | Disk | No | Revocation list | Self-certifying pathnames | Self-certifying pathnames, timestamps | Multiple distributed copies of RO files | 2 times slower than NFS for small files, comparable for files > 40 MB |
| **SNAD** | Disk | No | Lazy or active revocation options, no decision | Lockbox | Non-linear checksum of original text stored along with encrypted file | Potential DOS if lockbox server is compromised | Only the least secure option is comparable to a system w/o security |
| **PLUTUS** | Disk | No | Lazy revocation, revoked user retains same file permissions as time of revocation | Lockbox with user control over key dissemination. Users must secure distribution themselves | Stored encrypted, but requires augmentation to ensure integrity | Uses filegroups, but requires file owner to distribute keys | 1.4 times decrease from SFS for single file access |
| **SiRiUS** | Disk | No | Active Revocation | Combination of Master Encryption Key and File Encryption Keys | Freshness guarantee timestamp | Scalable to Internet, but requires file owner to distribute FEK | 20 times slower than NFS for small files, 2-6 times slower for 1MB files |

*Table 1 System Description*

Table 1 provides a summary of the characteristics described in the previous sections.

| Security Measure | Confidentiality | Integrity | Availability | Performance | Total |
|---|---|---|---|---|---|
| Encrypt-on-disk | 2 | 2 | 0 | 0 | 4 |
| Encrypt-on-wire | 1 | 1 | 0 | 0 | 2 |
| Threshold scheme | 1 | 1 | 2 | 2 | 6 |
| Timestamps | 0 | 2 | 0 | 2 | 4 |
| Digital signatures | 0 | 2 | 0 | 0 | 2 |
| Checksums | 0 | 1 | 0 | 1 | 2 |
| Lazy revocation | 1 | 0 | 2 | 2 | 5 |
| Active revocation | 2 | 1 | 1 | 0 | 4 |
| Key distribution server | 1 | 0 | 1 | 1 | 3 |
| Manual key distribution | 2 | 0 | 0 | 0 | 2 |
| Lockbox key mechanism | 2 | 0 | 1 | 1, 2* | 4,5* |
| Self-Certifying pathnames | 1 | 0 | 2 | 1 | 4 |
| Filegroups | 0 | 0 | 1 | 1 | 2 |
| Comprehensive versioning | 0 | 2 | 2 | 2 | 6 |

*Table 2 Quantitative Comparison of Security Measures*

The table uses a scale between 0-2 to assign values to the level of security of a particular security attribute. 0 denotes that the security measure does not affect the attribute, 1 denotes that the measure enhances security, 2 denotes that the measure significantly enhances security. For performance, the scale is modified to include three values (0-2) due to vastly differing performance costs. A 0 denotes that the measure significantly slows down system performance, a 1 denotes a significant performance penalty, and a 2 indicates a limited performance penalty (all systems suffer some performance degradation as a result of added security measures). While these figures are based on empirical data derived from the authors' research of each system, it is important to note that they are a subjective representation. *PLUTUS' key rotation scheme enhances performance.

| Security Measure | NASD | PASIS | S4 | CFS | SFS-RO | SNAD | PLUTUS | SiRiUs |
|---|---|---|---|---|---|---|---|---|
| Encrypt-on-disk | | | | X | X | X | X | X |
| Encrypt-on-wire | X | X | X | | | | | |
| Threshold scheme | | X | | | | | | |
| Timestamps | X | | | | X | | | X |
| Digital signatures | | | | | X | | X | |
| Checksums | X | | | | | X* | | |
| Lazy revocation | | | | | X | X | X | |
| Active revocation | X | | | X | | | | X |
| Key distribution server | X | | | | | | | |
| Manual key distribution | | | | X | | | | |
| Lockbox key mechanism | | | | | | X | X | |
| Self-certifying pathnames | | | | | X | | | |
| Filegroups | | | | X | | X | X | |
| Comprehensive versioning | | | X | | | | | |

*Table 3 Security Measure Applications to Surveyed Systems*
*\* Digital signatures are not considered for SNAD since the authors determined that they were too expensive*

Table 3 maps security measures to the systems that use them. An "X" denotes that the system applies the security measure in some form.

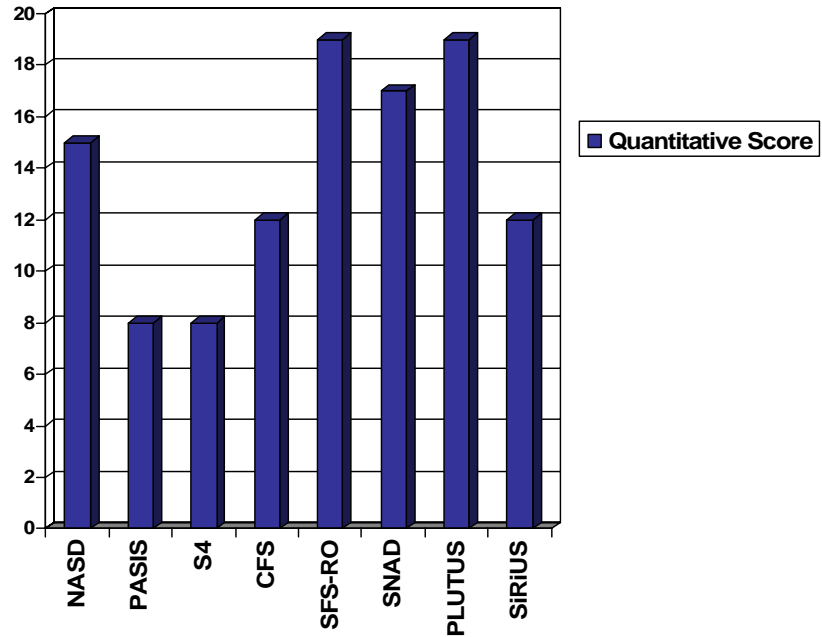

*Figure 4 Quantitative Comparison of Secure Storage Systems*

The quantitative values in Figure 4 are derived by summing the total scores of each security measure that applies to each system. While the numbers provide some measure for comparison, one must consider the purpose of each system in conjunction with the evaluation. For instance, both PASIS and S4 are survivable storage systems which should be used as part of a larger, secure storage system. Most authors mention that using their design in conjunction with a survivable storage system would provide the best security. Additionally, SFS-RO receives a high quantitative score, but it is limited to read-only applications.

## 5. CONCLUSIONS

All of the systems share a common goal: to protect stored data from the effects of a malicious adversary. From this common ground, however, the design approaches to reach this goal vary tremendously. Some systems aim to prevent an adversary from ever having access to data, while others assume that intrusions are inevitable and try to limit the amount of damage an intruder can introduce. Some systems separate data onto multiple storage servers to eliminate a single point of attack, and others rely on centralized trusted servers to effectively manage cryptographic keys. Some systems store encrypted data and others require encryption prior to transmitting messages on the wire. All of these examples present large fundamental differences that provide options to potential users of a storage security medium. It is very difficult to make direct comparisons between the systems because of the varied approaches, but potential users can select the most applicable solution to their specific problems.

The most secure solution will likely be a combination of the systems described. In fact, the majority of the designers of the systems recommend that their solution be part of a larger security plan. For example, if a user can accept additional cryptographic latency there is no reason to avoid encrypting data before applying a threshold scheme. The result would provide the security of encryption without relying on a trusted server and would increase the degree of availability. The problem with such layering, however, is the performance penalty. It is therefore a design requirement to analyze the tradeoffs between security and performance.

# REFERENCES

1. Mehmet Bakkaloglu, Jay J. Wylie, Chenxi Wang, Gregory R. Ganger. **On Correlated Failures in Survivable Storage Systems**. CMU SCS Technical Report CMU-CS- 02-129. May 2002.
2. Scott A. Banachowski, Zachary N. J. Peterson, Ethan L. Miller, and Scott A. Brandt. **Intra-file security for a distributed file system**. In *Proceedings of the 19th IEEE Symposium on Mass Storage Systems and Technologies*, pages 153–163, College Park, MD, April 2002. IEEE.
3. Matt Blaze, **A Cryptographic File System for Unix,** *First ACM Conference on Communications and Computing Security*, Fairfax, VA November, 1993.
4. Brian Cornell, Peter A. Dinda, Fabian E. Bustamante **Wayback: A User-level Versioning File System for Linux**. In *Usenix Annual Technical Conference*, Boston, MA Jun 27 – Jul 2, 2004
5. William Freeman and Ethan Miller. **Design for a decentralized security system for network-attached storage**. In *Proceedings of the 17th IEEE Symposium on Mass Storage Systems and Technologies*, pages 361–373, College Park, MD, March 2000.
6. Kevin E. Fu. **Group Sharing and Random Access in Cryptographic Storage File Systems.** Massachusetts Institute of Technology, Jun 1999. (Cepheus)
7. Kevin Fu, M. Frans Kaashoek, David Mazieres. **Fast and Secure Distributed Read Only File System.** In *Proceedings of the 4$^{th}$ USENIX Symposium on Operating Systems Design and Implementation,* pages 181-196, Sand Diego, CA, Oct 2000.
8. Gregory R. Ganger, Pradeep K. Khosla, Mehmet Bakkaloglu, Michael W. Bigrigg, Garth R. Goodson, Semih Oguz, Vijay Pandurangan, Craig A. N. Soules, John D. Strunk, Jay J. Wylie. **Survivable Storage Systems**. DARPA Information Survivability Conference and Exposition (Anaheim, CA, 12-14 June 2001), pages 184-195 vol 2. IEEE, 2001.
9. Garth Gibson, David Nagle, Khalil Amiri, Fay Chang, Howard Gobioff, Erik Riedel, David Rochberg, Jim Zelenka, "**Filesystems for Network-Attached Secure Disks**" *CMU Computer Science Technical Report, CMU-CS-97-118*. July 1997.
10. Howard Gobioff, David Nagle, Garth Gibson, "**Embedded Security for Network-Attached Storage**", *CMU SCS technical report CMU-CS-99-154,* June 1999.

11. Howard Gobioff, Garth Gibson, Doug Tygar, "**Security for Network Attached Storage Devices**", *CMU SCS technical report CMU-CS-97-185* 1997.
12. Eu-Jin Goh, Hovav Shacham, Nagendra Modadugu, and Dan Boneh**. SiRiUS: Securing Remote Untrusted Storage**. In the proceedings of the Internet Society (ISOC) Network and Distributed Systems Security (NDSS) Symposium 2003
13. Garth Goodson, Jay Wylie, Greg Ganger & Mike Reiter. **Decentralized Storage Consistency via Versioning Servers.** Carnegie Mellon University Technical Report CMU-CS-02-180, September 2002.
14. Garth R. Goodson, Jay J. Wylie, Gregory R. Ganger, Michael K. Reiter. **Efficient Consistency for Erasure-coded Data via Versioning Servers**. Carnegie Mellon University Technical Report CMU-CS-03-127, April 2003.
15. Garth R. Goodson, Jay J. Wylie, Gregory R. Ganger, Michael K. Reiter. **A Protocol Family for Versatile Survivable Storage Infrastructures.** Carnegie Mellon University Parallel Data Lab Technical Report CMU-PDL-03-104, December 2003.
16. Mahesh Kallahalla, Erik Riedel,Ram Swaminathan, Qian Wang, and Kevin Fu. **PLUTUS: Scalable secure file sharing on untrusted storage**. In Conference on File andStorage Technology (FAST'03) pp. 29-42 (31 Mar - 2 Apr 2003, San Francisco, CA). Published by USENIX, Berkeley, CA.
17. David Mazieres, Michael Kaminsky, M. Frans Kaashoek, Emmett Witchell. **Separating Key Management From File System Security.** In 17[th] ACM Symposium on Operating Systems Principles, pages 124-139, Dec 1999.
18. David Mazieres, Dennis Shasha **Building Secure File Systems Out of Byzantine Storage.** In *Proceedings of the Twenty-first Annual Symposium on Principles of Distributed Computing*, pages108-117, Monterey, CA, 2002.
19. David Mazieres, Dennis Shasha. **Don't Trust Your File Server.** In Proceedings of the 8[th] Workshop on Hot Topics in Operating Systems, 2001.
20. Ethan L. Miller, Darrell D. E. Long, William Freeman, and Benjamin Reed. **Strong security for distributed file systems**. In *Proceedings of the 20th IEEE International Performance, Computing and Communications Conference (IPCCC '01)*, pages 34–40, Phoenix, April 2001. IEEE.
21. Ethan L. Miller, Darrell D. E. Long, William E. Freeman, and Benjamin C. Reed. **Strong security for network-attached storage**. In *Proceedings of the 2002 Conference on File and Storage Technologies (FAST)*, pages 1–13, Monterey, CA, January 2002.
22. Kiran-Kumar Muniswamy-Reddy, Charles P. Wright, Andrew Himmer, Erez Zadok **A Versatile and User-Oriented Versioning File System.** In *3[rd] Usenix Conference on File and Storage Technologies* (FAST 2004)
23. Adam Pennington, John Strunk, John Griffin, Craig Soules, Garth Goodson & Greg Ganger. **Storage-based Intrusion Detection: Watching Storage Activity For Suspicious Behavior.** 12th USENIX Security Symposium, Washington, D.C., Aug 4-8, 2003. Also available as Carnegie Mellon University Technical Report CMU-CS-02-179, September 2002.
24. Benjamin C. Reed, Edward G. Chron, Randal C. Burns, and Darrell D. E. Long. **Authenticating network-attached storage**. *IEEE Micro*, 20(1):49–57, January 2000.

25. Erik Riedel, Mahesh Kallahalla, and Ram Swaminathan. **A Framework for Evaluating Storage System Security.** In *Proceedings of the 1<sup>st</sup> Conference on File and Storage Technologies*, Monterey, CA, Jan 2002.
26. Craig A.N. Soules, Garth R. Goodson, John D. Strunk, Gregory R. Ganger **Metadata Efficiency in Versioning File Systems.** In *2<sup>nd</sup> USENIX Conference on File and Storage Technologies*, San Francisco, CA mar 31-Apr 2, 2003
27. John D. Strunk, Garth R. Goodson, Adam G. Pennington, Craig A.N. Soules, Gregory R. Ganger. **Intrusion Detection, Diagnosis, and Recovery with Self-Securing Storage**. CMU SCS Technical Report CMU-CS-02-140, May 2002.
28. John D. Strunk, Garth R. Goodson, Michael L. Sheinholtz, Craig A.N. Soules, Gregory R. Ganger, **Self-Securing Storage: Protecting Data in Compromised Systems**. In *4<sup>th</sup> Symposium on Operating System Design and Implementation*, San Diego, CA Oct 2000
29. Theodore Ming-Tao Wong. **Decentralized Recovery for Survivable Storage Systems.** Carnegie Mellon School of Computer Science Ph.D. Dissertation CMU-CS-04-119. May 2004.
30. Charles P. Wright, Jay Dave, Erez Zadok. **Cryptographic File Systems Performance: What You Don't Know Can Hurt You.** In *IEEE Security in Storage Workshop (SISW 2003),* Oct 2003.
31. Charles P. Wright, Michael C. Martino, Erez Zadok. **NCryptfs: A Secure and Convenient Cryptographic File System.** In *USENIX 2003 Annual Technical Conference,* Jun 2003.
32. Jay J. Wylie, Mehmet Bakkaloglu, Vijay Pandurangan, Michael W. Bigrigg, Semih Oguz, Ken Tew, Cory Williams, Gregory R. Ganger, and Pradeep K. Khosla. **Selecting the right data distribution scheme for a survivable storage system.**. Technical report CMU-CS-01-120. Carnegie Mellon University, May 2001.
33. Jay J. Wylie, Michael W. Bigrigg, John D. Strunk, Gregory R. Ganger, Han Kiliccote, and Pradeep K. Khosla. **Survivable information storage systems.** IEEE Computer, 33(8):61-68, August 2000.
34. Erez Zadok. **Stackable File Systems as a Security Tool.** Columbia University CS TechReport CUCS-036-99, Dec 1999.